# On the Dynamics of Local to Global Campaigns for Curbing Gender-based Violence

Prakruthi Karuna[1], Hemant Purohit[1], Bonnie Stabile,[2], Angela Hattery[3]

[1]Information Sciences and Technology, George Mason University, VA, USA
[2]School of Policy, Government, and International Affairs, George Mason University, VA, USA
[3]Women and Gender Studies, George Mason University, VA, USA

{pkaruna,hpurohit,bstabile,ahattery}@gmu.edu

**Abstract**. Gender-based violence (GBV) is a human-generated crisis, existing in various forms, including offline, via physical and sexual violence, and now online via harassment and trolling. While studying social media campaigns for different domains such as public health, natural crises, etc. has received attention in the literature, such studies for GBV are still in nascent form. The dynamics of campaigns responding to curb this crisis could benefit from systematic investigation. To our knowledge, this is the first study to examine such public campaigns involving social media by organizations operating at the local, national and global levels, with an eye to answering the following research questions: (1) How do members of one campaign community engage with other campaign communities? (2) How do demographic variables such as gender effect campaign engagement in light of given regional crime statistics? (3) Is there any coordination among organizational users for campaigns with similar underlying social causes?

## 1 Introduction

Gender-based violence (GBV) encompasses "acts of violence ranging from online harassment to domestic assault and human trafficking" [1]. "This violence, and its ongoing threat, interferes with every major capability in a woman's life" [4] with clear consequences for physical and emotional well-being, compromising the ability to engage fully and be equitably represented in governmental and economic spheres of power.

Unfortunately, it appears as though GBV knows no geographical or demographic bounds. According to the United Nations (UN) entity for gender equality and the empowerment of women—UN Women [5], globally "1in 3 women experience physical or sexual violence, mostly by an intimate partner". In the United States (US) alone, it is estimated that nearly "1 in 5 women are raped", and "1 in 4 women experience severe physical violence by an intimate partner" at some point during their lifetimes [6]. The American Association of Universities' Campus Climate Survey reports a similar incidence of sexual assault - 23 percent - at institutions of higher education in the US [7].

Organizations at levels ranging from the local to the global have arisen in various forms to combat the problem of GBV. In the social media context, we will investigate the impact of following three anti-GBV campaign communities (*a campaign community consists of social media users, who use the campaign's identity hashtag, in the shared messages*):

1. *(College-centric)* The US White House **#ItsOnUs** initiative [2,8], to combat sexual assault on college campuses in the US;
2. *(Nation-centric)* The US White House **#StateOfWomen** initiative [3,9], one facet of which focuses on Violence Against Women in the US; and
3. *(Globe-centric)* The UN **#HeForShe** initiative [1,10], which seeks to involve men and boys in the cause of establishing gender equality for women, including ensuring that women and girls can live in environments free from the threat of violence.

Our ultimate goal is to understand what shapes people's engagement towards campaigns designed for responding to the human-generated crisis of GBV, with focus on understanding the reach of such campaigns in social media in the U.S. region.

**Research Questions.** Given the adoption of the Twitter platform as a major vehicle for anti-GBV campaigns by social welfare organizations, we study the following research questions on Twitter for the three campaigns (*#ItsOnUs*, *#StateOfWomen*, *#HeForShe)*:

1. How do members of one anti-GBV campaign community engage with other anti-GBV campaign communities on Twitter?
2. How do demographic variables such as gender effect the campaign community engagement in light of given regional crime statistics?

3. Is there any coordination among organizational users engaging in different campaigns with similar underlying social causes?

This paper is organized in the following as related work in section 2, data collection and processing method in section 3, followed by research question analyses in section 4, and discussion in section 5 before conclusion.

## 2 Related Work

We present related works for studying anti-GBV campaign dynamics along sociological analysis of engagement in the anti-GBV activism, policy impact of understanding the engagement dynamics, and relevant computational social science research on Twitter.

**Sociological Analyses of anti-GBV activism**. While engagement by genders in social media in general has been reported to be biased to types of social media, there is a likelihood of different dynamics for awareness campaigns, especially for genders by regions of high versus low crime rates. Research survey reported in the Huffington Post [11] reveals that though men and women both use various forms of social media, women comprise 62% of Twitter users. Not surprisingly, younger people are far more likely to use Twitter than middle-aged and older adults. Thus, when we consider men's and women's involvement in anti-GBV social activism, it is not surprising that women are significantly more likely to be involved, at every level from social media campaigns, to activism, to advocacy than are men [12,13]. Much attention has focused recently on programs designed to encourage men's engagement in anti-GBV activism, specifically sexual assault, including initiatives such as Coaching Men Into Boys, and "Take the Pledge Violence Against Women: We Can Stop It"—a program developed by the US Attorney General's Office [14]. One interesting transnational example is seen in the use of social media by women in the social movement focused on reducing GBV as part of the larger Arab Spring [15]. We hypothesize that women, who are both more likely to use Twitter and more likely to be engaged in anti-GBV activism, will use social media in response to the three campaigns we are examining. Feminist activism has long been focused on collaboration and grassroots activism, and the anti-violence movements in the US that began in the early 1970s were no different. Early anti-violence activists understood clearly the connections between GBV and other forms of injustice against women, including unequal pay and reproductive justice [16]. Thus, we would expect that there will be evidence to support similar levels of collaboration among active members of the various campaign communities we research in this study and beyond.

**Policy analyses of engagement dynamics**. The three campaigns that we examine represent efforts by governmental or intergovernmental organizations to leverage social actors to agitate for desired social and policy change. Prior to the advent of current social media platforms - recall that Twitter was founded in 2006 - political scientists had provided evidence that networks such as these, of state and nonstate actors including civil societies and international organizations, could be influential in issue definition and agenda setting, and influencing policy change in states, international organizations and among private actors [17]. Such networks "are bound together by shared values, a common discourse, and dense exchanges of information" [17]. The *#ItsOnUs* website [2] features a section highlighting groups and organizations that act as partners in its efforts to address the problem of sexual assault, an explicit acknowledgement of the importance of coalition building in advancing social and cultural change. The impetus for stakeholders to reject the status quo and advocate for policy and social change can also come from public figures identified as "policy entrepreneurs" by Kingdon [18], who are prominent public figures invested in raising awareness of issues by parlaying their fame into effective spokesmanship for a cause. The three campaigns of interest here all have such entrepreneurs associated with them, whose names may emerge through the analysis. Punctuated equilibrium theory [19] might also suggest opportunities for raising awareness, through events, whether orchestrated (such as the 'State of Women' Summit [9]) or otherwise, that create new perceptions around policy issues. The present analysis can help shed some light on how and to what extent certain anti-GBV individual and organizational actors engage and coordinate in seeking to effect change for a common social cause.

**Computing analyses for GBV on Social Media**. Prior literature on the computational social science efforts to study GBV domain on social media present different explorations on the dimensions of forms, language use, and content analysis. While most of the prior research has focused largely on the forms of cyberbullying and online harassment/trolling [20,21], few studies have been performed to examine connection between public attitudes and GBV issues by means of sharing voices and engaging on social media. Among the regionally focused studies, [22] showed a relationship between misogynistic language and rape statistics in the U.S., the most infamous component of GBV, which motivates our second research question on studying campaign effectiveness by region. [23] focused on the U.K. region to study engagement of the general population by location, gender, and language characteristics. Among the globally focused studies, [24] provided a comprehensive large-scale study of Twitter messages, by analyzing a broader topical dataset for different forms of GBV—physical violence, sexual violence, and harmful socio-cultural practices, with respect to language usage beyond just misogyny, gender-wise participation, and content practices. The present analysis complements prior studies focused on public attitudes towards GBV by understanding the dynamics of user engagement in the anti-GBV campaign communities based on user interaction networks, user location, types (individual and organization), and gender.

**Table 1. Dataset Statistics for crawling period: June 14 - June 21, 2016. Sets of various combinations of campaign related tweets are mutually exclusive.**

|  | Total Tweets | Retweets (% of Total) | Total Authors |
| --- | --- | --- | --- |
| All containing any of campaign hashtags | 168,950 | 124,952 (74%) | 72,957 |
| #ItsOnUs | 1,415 | 1,048 (74%) | 1,095 |
| #StateOfWomen | 157,288 | 115,467 (73%) | 66,615 |
| #HeForShe | 9,112 | 7,564 (83%) | 6,452 |
| #ItsOnUs & #StateOfWomen | 890 | 734 (82%) | 656 |
| #ItsOnUs & #HeForShe | 5 | 1 (20%) | 5 |
| #StateOfWomen & #HeForShe | 239 | 138 (58%) | 97 |
| #ItsOnUs & #StateOfWomen & #HeForShe | 1 | 0 (0%) | 1 |

## 3 Data Collection and Processing Method

We first discuss our approach to data collection, followed by the extraction process for various user metadata, including the user type (individual versus organization), individual user gender, and user location from the Twitter profile.

### 3.1. Data Collection

We adopt a keyword-based crawling method for collecting data from Twitter, which is the most common method for Twitter data studies in the prior literature. For collecting Twitter data related to a given set of anti-GBV campaigns, we have a three step process. First, we prepare a seed set of keywords relevant to the campaigns. Second, we use the Twitter Streaming API[1], 'filter/track' method, which provides a stream of public tweets containing any of the seed keywords. Third, we extract and store all the relevant metadata such as tweet text, timestamp of posting, tweet type such as retweets, authoring user screen names, and its self-reported author profile information such as full name, and location.

Our seed set of keywords for crawling includes three hashtags corresponding to each of the three diverse focus campaigns - *#StateOfWomen* (Nation-centric), *#HeForShe* (Globe-centric), and *#ItsOnUs* (College-centric). We collected the data for one week, after the announcement of the *#StateOfWomen* campaign, starting from June 14 to June 21, 2016. *#StateOfWomen* initiative launching was a timely opportunity to observe the engagement of users on social media across different campaigns with similar underlying social cause. Table 1 provides the basic statistics of the dataset.

### 3.2. Interaction Network Creation

We create who-talks-to-whom interaction networks for all the three campaign communities, where a node represents a user and an edge represents an interaction between two users. Given two user nodes A and B, a directional edge is created from A to B if '*A retweets or mentions or replies B*', and the edge weight is assigned by the weighted sum of frequency of the specific interaction (retweeting another user's post, or mention of another user in a post, or reply to another user's post).

---

[1] Twitter Streaming API: http://dev.twitter.com/streaming/reference/post/statuses/filter

For an interaction network $G$ constructed from a corpus of Twitter posts $P$ for the campaign(s), we define $G = <V,E>$ as a directed network of a set $V$ of $n$ user nodes and a set $E$ of $m$ directed edges, where the tie strength of an edge $E_{AB}$ is defined as:

$$Strength(E_{AB}) = \alpha * \sum_{1 \leq i \leq |P|} Retweets_i(A,B) + \beta * \sum_{1 \leq j \leq |P|} Mentions_j(A,B)$$
$$+ \gamma * \sum_{1 \leq k \leq |P|} Replies_k(A,B) \quad \text{.. (1)}$$

s.t.
$$A, B \in V; \quad i, j, k \in \mathbb{N}; \quad 0 \leq \alpha, \beta, \gamma \leq 1$$

$Retweets_i(A,B) = 1$ if $A$ forwards a post of $B$ in the post $ith$, else 0
$Mentions_j(A,B) = 1$ if $A$ mentions $B$ in the post $jth$, else 0
$Replies_k(A,B) = 1$ if $A$ replies to a post of $B$ in the post $kth$, else 0

We compute various network characteristics such as the node degree, and clustering coefficient, by using Gephi[2] open source platform, and explore the generated interaction networks for different user attributes (e.g., gender), discussed in section 4.

### 3.3. User Metadata Extraction

Understanding engagement in the campaign communities via interaction networks requires a systematic approach to understand the community members, the users. Therefore, we extract various user metadata as the following.

***User Type: Individual vs. Organization***. Researchers in computational social science have investigated various methods to classify user types on Twitter, such as based on demographic attributes, influence, and ideologies. This study considers the user type of represented identity in the user profile—organization versus individual, to understand dynamics of organizational engagement in anti-GBV campaigns. An organizational type of user account is the one that represents a group, company or an organization, such as a university's account @GeorgeMasonU, or a NGO account @EndRapeOnCampus on Twitter. All the other users are considered individual accounts. The scope of this paper is limited to these two user type classes, and we plan to extend this in future.

We adopt the approach of [25] for individual versus organization type classification, which provides an implementation on GitHub[3]. It takes input as the tweet json objects including tweet and user information as returned by the Twitter API in our data collection. In a manual verification study of randomly selected 50 users from the dataset, we found 8 inconsistent predictions for the adopted classifier, giving an accuracy of 84%. We expect that organizational users representing group identity are likely to be lesser in general, and table 2 reflects such a distribution by user type.

---

[2] https://gephi.org/

**Table 2: User Type Distribution.**

|                | Users  | % of Total | Tweets  | % of Total |
|----------------|--------|------------|---------|------------|
| Total          | 72,957 |            | 168,950 |            |
| Organizational | 86     | 0.11%      | 8,574   | 5%         |
| Individual     | 72,871 | 99.88%     | 160,376 | 94.9%      |

*User Gender: Male vs. female vs. Unisex.* For the gender classification process, we only consider users that were classified as individuals in the preceding step. Twitter does not ask users to provide gender in the registration process, and therefore, several studies have proposed methods to classify a user gender on Twitter [26,27], using a diverse set of features such as user name, user-generated content, user profile, profile images, etc. Gender classification is challenged by the dynamic nature of content style and writing practices, as well as the self reported nature of the user profile data.

**Table 3: Gender-wise distribution of Individual Users.**

|       | Individual Users | % of Total | Tweets  | % of Total |
|-------|------------------|------------|---------|------------|
| Total | 72,871           |            | 160,376 |            |
| Male  | 13,826           | 18.97%     | 25,045  | 15.6%      |

---

[3] https://github.com/networkdynamics/humanizr

| | | | | |
|---|---|---|---|---|
| Female | 30,733 | 42.2% | 69,289 | 43.2% |
| Unisex | 1,302 | 1.8% | 2,895 | 1.8% |
| Unknown | 27,010 | 37.1% | 63,147 | 39.4% |

The existing implementations from the prior studies using content and profile features were not found to be efficient in our manual verification study of randomly selected 50 users, such as maximum accuracy of 34% while using only users tweets in GBV campaign data) and 46% when using user's previous tweets for [27]. Therefore, we resolved to user name based gender identification, on the expense of poor dataset coverage though, using input as user profile data for the real name field whenever available. This tool is available on GitHub[4] and provides reference to the name database for each country, and also the database of first names from all around the world provided together with gender.c—an open source C program for name-based gender inference. For our manual verification study of randomly selected 50 users, we found accuracy of 78%. Table 3 shows statistics by gender, and we note the consistency with literature on the higher female GBV activism.

*User Location: Focus on US region.* We use the author profile location field for a user to study the geographical engagement in the campaigns across US. We map profile location metadata to the US states by querying the textual location field values in open source search tool Nominatim based on OpenStreetMap data[5]. For the 47% of users, we resolved the location metadata given the noisy, but creative user-generated location field values, such as '*IN THE FREE MARKET*'. We use user locations by states to study the patterns of state level campaign engagement, and the reported hate-crime statistics in US in 2014 (selected categories of interest in the data: *Gender, Gender Identity*)—available through FBI Uniform Crime Reporting (UCR) Program[6].

---

[4] https://github.com/tue-mdse/genderComputer
[5] https://github.com/twain47/Nominatim
[6] https://ucr.fbi.gov/about-us/cjis/ucr/hate-crime/2014/topic-pages/jurisdiction_final

## 4 Research Question Analysis

We use the method described in section 3 to generate various user metadata for the members of different campaign communities, and analyze them.

### 4.1. RQ1 - Crossover of User Community Engagement

For finding overlapping engagement by the users in the different campaign communities, we compute Jaccard Similarity Coefficient, which measures similarity between finite sample sets, and is defined as the size of the intersection divided by the size of the union of the sample sets. For instance, for users participating in two campaign communities *#ItsOnUs* and *#StateOfWomen*, we define the *Overlap* measure as:

$$Overlap(\#ItsOnUs, \#StateOfWomen) = \frac{\{\text{Users in } \#ItsOnUs\} \cap \{\text{Users in } \#StateOfWomen\}}{\{\text{Users in } \#ItsOnUs\} \cup \{\text{Users in } \#StateOfWomen\}}$$

.. (2)

**Table 4. Campaign Community *Overlap* of users.**

| Community A | Community B | Jaccard Coefficient | |
|---|---|---|---|
| | | All Users | Organizational Users |
| *#ItsOnUs* | *#StateOfWomen* | 0.02 | 0.49 |
| *#ItsOnUs* | *#HeForShe* | 0.01 | 0.15 |
| *#StateOfWomen* | *#HeForShe* | 0.01 | 0.13 |

Results in table 4 show a weak *overlap* between the members of the campaign communities, especially with the globe-centric campaign, despite having related social cause awareness. Engaged organizations appear to be more aware and connected in the case of college and nation-centric campaigns, partly also due to their common supporting user base—both are White House launched initiatives.

For studying the engagement of users in spreading campaign awareness by cross-referencing campaign identity hashtags in tweets, we create 7 subsets of the data as the following: Tweets containing a.) only *#ItsOnUs*, b.) only *#StateOfWomen*, c.) only #HeForShe, d.) both #ItsOnUs and *#StateOfWomen*, e.) both #ItsOnUs and *#HeForShe*, f.) both *#StateOfWomen* and *#HeForShe*, and g.) tweets containing all the three *#ItsOnUs*,

*#StateOfWomen* & *#HeForShe*. Table 1 shows the results for the tweet volumes under these 7 sets. Interestingly, there is only a user and a tweet which actually intersected all the three campaigns in a week-long period. Along the same line, although the cross-referencing for college and nation-centric campaigns *#ItsOnUs* and *#StateOfWomen* exists (albeit less than 10% compared to individual campaign tweet volumes), there is not much intersection with the globe-centric campaign.

Figure 1 shows the tweet content generation for the different campaigns by the gender. We noticed that college-centric campaign *#ItsOnUs* has a similar pattern with *#StateOfWomen* while not as much with *#HeForShe*, partly due to the higher community *overlap* as observed in table 4. Although we note the consistent pattern of higher female engagement as observed in the literature (discussed in section 2), we can note the higher proportions of male engagement in the *#HeForShe* by the very characteristic of the campaign cause. This in turn provides an insight on how campaigns could benefit by actually coordinating together to cross-referencing content strategies for improving engagement in specific demographics.

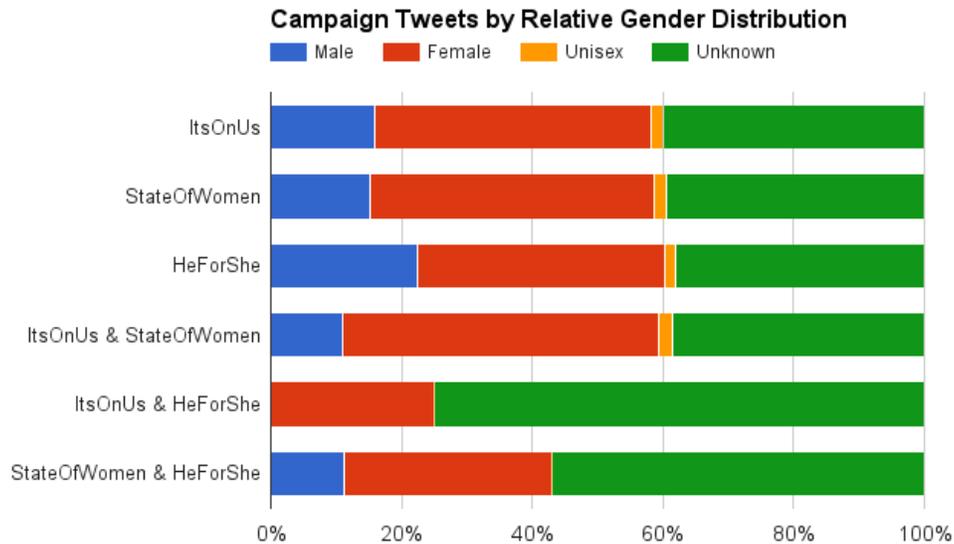

**Fig. 1. Gender-wise relative distribution of tweets in the three campaign communities, and their intersections.**

### 4.2. RQ2 - Regional Overlap of demographic data with Reported Crime Data

We computed correlation between the state-wise statistics of GBV related crime volumes as discussed earlier in section 3.3, and the volumes of users participating from those regions to measure the regional engagement pattern. We noticed a positive correlation coefficient of 0.34, indicating Twitter engagement in consistence with the regional crime rates. Figure 2 shows the comparison of state-wise distribution. While we noticed that majority of states have high Twitter engagement, there is an opposite trend for some states like South Dakota, indicating the need for region-aware engagement strategies.

For gender-specific patterns across the states, we did not observe any different pattern than the literature on offline GBV activism—female participation was observed higher than the male participation across states and across campaigns, indicating a greater need to increase the engagement of male and unisex users for the campaign design.

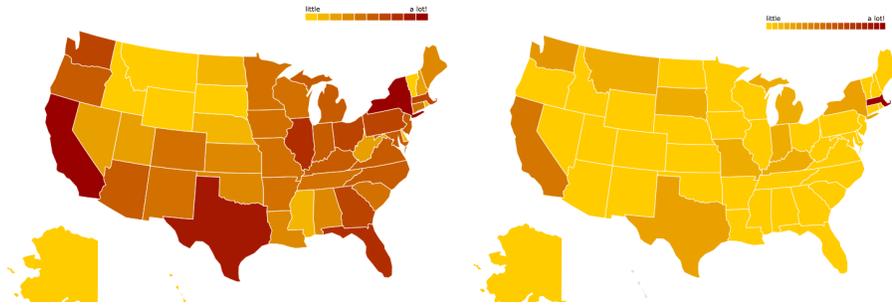

**Fig. 2. Comparison of different U.S. states by user engagement in anti-GBV campaigns (a), and the 2014 GBV related statistics reported in the FBI UCR data. Darker color represents higher intensity value.**

### 4.3. RQ3 - Organizational User Interaction

We created interaction network for users in all three campaign communities as well as overall (we only present this one in figure 3 and 4 owing to space limitation and having any interesting patterns). We chose the     and     weights to be equal for *Replies* and *Mentions* functions, and double the value of     for *Retweets*, given they represent a deeper engagement of conversation between two users than forwarding a message.

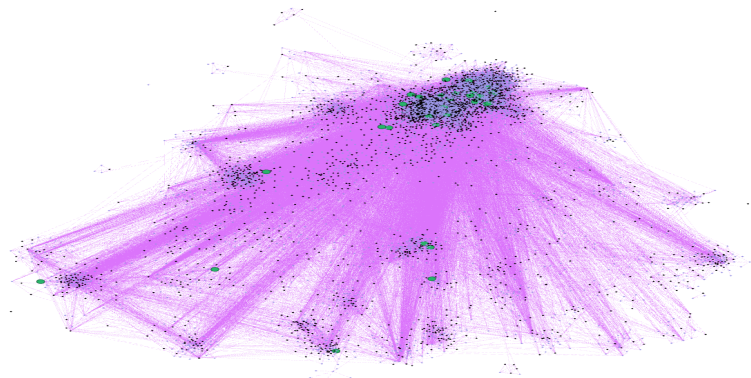

**Fig. 3. Individual users (black) and organizations (Green) in the interaction network.**

The interaction network is overall dense and with low modularity, e.g., the giant component of the interaction network of all campaigns together has more than 35,000 users even with minimum out degree 2 (i.e., a user has at least engaged twice in writing/sharing a message). Therefore, to get better insights about our key goal of understanding higher engagement patterns, we chose to limit the interaction network of all campaign users to minimum out degree 5, and analyzed the giant component by user metadata of type (organization and individual) in figure 3 and gender (male and female) in figure 4. A strong interaction dynamics can be observed in figure 3 for organizational users (green nodes) and individual users (black nodes) with higher average values of clustering coefficient, degree, and connected components than individuals alone. Figure 4 shows higher interaction and coordination between female users (pink nodes) than the male users (blue nodes), who are sparsely connected compared to females.

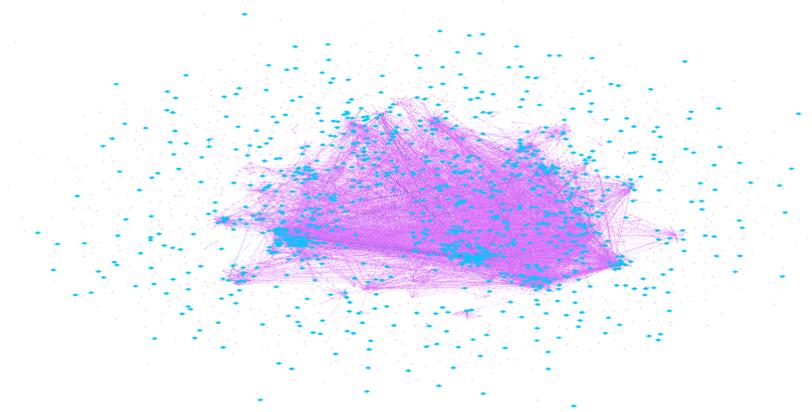

**Fig. 4.** Male (Blue) and Female (Pink) users in the interaction network.

## 5 Discussion

We first discuss the limitations, followed by lessons, and lastly future directions.

*Limitations.* The presented analyses relied on existing user attribute extractors, such as gender and organization classifiers, which have accuracy limitations, and could be biased. User location mapping was limited to profile location metadata, which may be unavailable sometimes, and other techniques for location predictions based on profile tweets could be an alternative. We also note that the reported crime statistics are often under-reported due to the sensitive nature of the GBV related crimes [5], however, patterns across the states is less like to change.

*Lessons Learned.* In the Violence Against Women movement that crystallized later in the 20th century, activist partners had "learned a strategic lesson 'Let's look for more allies' "; this was done, in part by linking the call for women's rights to the wider issue of human rights overall [17]. Such a lesson and the observations from the presented analyses inform about the need of better coordination among campaign strategists for framing of the issue in social media messaging, such that users can be engaged efficiently across the campaign levels in not just sharing but actually curating the content. For example, *#StateOfWomen* has a larger scope of women's rights than GBV that could be leveraged to design campaign outreach for engaging a diverse, bigger set of stakeholders, and other campaign communities like college-centric *#ItsOnUs* could benefit by coordinating with the larger scope campaign community in spreading awareness via cross-referencing.

*Future Work.* We plan to study user engagement by further extending the demographic attributes such as age and race, as well as modeling relationship metadata from the user profiles—husbands and wives. Another future direction is to investigate the content, language of campaigners that actually motivate user engagement for deeper mode of content curation beyond information sharing, e.g., in table 1, tweets including both *#StateOfWomen* and *#HeForShe* had lesser retweets (58%) and more original posts than other cases. We will explore the latent factors that lead to such user actions.

## 6 Conclusion

We presented a first study of user engagement in anti-GBV campaign communities on social media, by analyzing three diverse—college/nation/globe-centric campaigns on Twitter via cross-community participation of genders, and organization user type as well as comparison of regional crime rates. We observed a consistent pattern of higher female participation in anti-GBV activism likewise offline, while the target demographic of males lack engagement, and the organizational users lack coordination for effective outreach. This study will help design better practices of interrelated social media campaigns.

### Acknowledgement

Authors are grateful of "It's On Us" campaign team for collaboration, and providing a key motivation for this study to leverage research analyses into social impact work.

## References

1. "HeForShe". (2016) Violence. UN Women.Org. Available at http://goo.gl/QZ2baP.
2. "ItsOnUs". (2016). It's On Us Campaign. Available at http://goo.gl/xponHJ.
3. "StateOfWomen". (2016). The United States Of Women Campaign. Available at http://goo.gl/JsxwpL
4. Nussbaum, Martha (2005). Women's Bodies: Violence, Security, Capabilities. *Journal of Human Development,* 6 (2): 167-183.

5. UN Women. (2016). Facts and Figures: Ending Violence Against Women. Available at http://goo.gl/V0Ur6o
6. Centers for Disease Control and Prevention (2010) National Intimate Partner Sexual Violence Survey: 2010 Summary Report. Available at http://goo.gl/DxTAqW
7. American Association of Universities (2015). Campus Climate Survey on Sexual Assault and Sexual Misconduct. Available at https://goo.gl/ABuvzZ
8. White House. (2014). President Obama Launches the "It's On Us" Campaign to End Sexual Assault on Campus. September 19. Available at https://goo.gl/QSP0lY
9. White House. (2016). "Together, We Are Stronger." June 6. Available at https://goo.gl/mOeW82
10. UN News Centre (2014). Ahead of International Women's Day, UN Asks Men to 'Stand Up and Deliver' on Human Rights for All. Available at http://goo.gl/jmM1Ln
11. Huffington Post. (2012). "Social Media By Gender: Women Dominate Pinterest, Twitter, Men Dominate Reddit, YouTube (INFOGRAPHIC)." *Huffington Post*. June 21, 2012. http://goo.gl/jcMCb6
12. Messner, Michael A. (2016). "BAD MEN, GOOD MEN, BYSTANDERS: Who Is the Rapist?" *Gender and Society*, 30 (1): 57-66.
13. Messner, Michael A., Max A. Greenberg, and Tal Peretz. (2015). Some men: Feminist allies and the movement to end violence against women. New York: Oxford University Press.
14. National Resource Center on Domestic Violence. (2011). *Men in the Movement to End Violence against Women: Campaigns and Campaign Materials*. Web. 13 July 2016.http://goo.gl/sY7C3V
15. Arab Social Media Report. November (2011). The Role of Social Media in Arab Women's Empowerment. Available at http://goo.gl/iEHT4H
16. Renzetti, Claire M., Jeffery L. Edleson, and Raquel Kennedy Bergen. 2001. *Sourcebook on Violence Against Women*. Thousand Oaks, CA: Sage.
17. Keck, Margaret E. and Kathryn Sikkink. 1998. *Activists Beyond Borders: Advocacy Networks in International Politics*. Ithaca: Cornell University Press.
18. Kingdon, John W. 2002. *Agendas, Alternatives and Public Policies*. Boston: Little, Brown and Company.

19. Baumgartner, Frank and Brian Jones (2009). *Agendas and Instability in American Politics.* Chicago: University of Chicago Press.
20. Coles, B. A., & West, M. (2016). Trolling the trolls: Online forum users constructions of the nature and properties of trolling. *Computers in Human Behavior*, *60*, 233-244.
21. Hosseinmardi, H., Mattson, S. A., Rafiq, R. I., Han, R., Lv, Q., & Mishra, S. (2015, December). Analyzing Labeled Cyberbullying Incidents on the Instagram Social Network. In *SocInfo'15,* 49-66. Springer.
22. Fulper, R., Ciampaglia, G. L., Ferrara, E., Ahn, Y., Flammini, A., Menczer, F., Lewis, B., & Rowe, K. (2014). Misogynistic language on Twitter and sexual violence. In *Proceedings of the ACM Web Science Workshop on Computational Approaches to Social Modeling (ChASM)*.
23. Bartlett, J., Norrie, R., Patel, S., Rumpel, R., & Wibberley, S. (2014). Misogyny on twitter. *Demos*. Available at http://goo.gl/x8OTTF
24. Purohit, H., Banerjee, T., Hampton, A., Shalin, V. L., Bhandutia, N., & Sheth, A. (2016). Gender-based violence in 140 characters or fewer: A #BigData case study of Twitter. *First Monday*, *21*(1).
25. McCorriston, J., Jurgens, D., & Ruths, D. (2015, April). Organizations Are Users Too: Characterizing and Detecting the Presence of Organizations on Twitter. In *Ninth International AAAI Conference on Web and Social Media*.
26. Pennacchiotti, M., & Popescu, A. M. (2011). A Machine Learning Approach to Twitter User Classification. *ICWSM*, *11*(1), 281-288.
27. Volkova, S., Coppersmith, G., & Van Durme, B. (2014). Inferring User Political Preferences from Streaming Communications. In *ACL (1)* (pp. 186-196).